\documentstyle[twocolumn,aps,epsfig,floats]{revtex}  

\begin{document} 

\newcommand{\be}{\begin{equation}}
\newcommand{\ee}{\end{equation}}
\newcommand{\bea}{\begin{eqnarray}}
\newcommand{\eea}{\end{eqnarray}}

\draft  

\twocolumn[\hsize\textwidth\columnwidth\hsize\csname %
@twocolumnfalse\endcsname

\title{
Instability and Spatiotemporal Dynamics of Alternans in Paced Cardiac Tissue   
}
\author{Blas Echebarria and Alain Karma} 
\address{Department of Physics and Center for 
Interdisciplinary Research on 
Complex Systems,\\
Northeastern University, Boston, MA 02115}

\date{November 23, 2001}

\maketitle

\begin{abstract}
We derive an equation that 
governs the spatiotemporal dynamics of small amplitude
alternans in paced cardiac tissue.
We show that a pattern-forming linear instability leads to the spontaneous
formation of stationary or traveling waves  
whose nodes divide the tissue into regions 
with opposite phase
of oscillation of action potential duration.
This instability is important because it creates 
dynamically an heterogeneous electrical substrate 
for inducing fibrillation if the tissue size exceeds a fraction of the
pattern wavelength. We compute this wavelength analytically 
as a function of three basic length scales characterizing
dispersion and inter-cellular electrical coupling.   
\end{abstract}

]

It is well established that the duration of
cardiac excitation can oscillate
from beat to beat at sufficiently short pacing interval \cite{Min}.
Pioneering studies by Nolasco and Dahlen \cite{NolDal}
and Guevara {\it et al.} \cite{Gueetal} have demonstrated that
the generic sequence LSLS... of long and short action
potential duration ($APD$), known as alternans, 
is a direct consequence of the restitution relationship
\begin{equation}
APD^{n+1}=f(DI^n)\label{res}
\end{equation}
between the $APD$ generated by the $n^{th}+1$ 
stimulus, $APD^{n+1}$, and the diastolic time interval
$DI^n$ during which the
tissue recovers its resting properties after
the end of the previous ($n^{th}$) action potential.
If we denote the interval   
between the $n^{th}$ and $n^{th}+1$ stimulus by $T^n$,
we must have $DI^n= T^n-APD^n$. 
Then, for a fixed period: $T^n=\tau$ 
for all $n$, Eq. \ref{res}  
yields the map $APD^n=f(\tau-APD^{n-1})$ that has
a period doubling instability if the 
slope $f'$ of the restitution curve evaluated at its
fixed point exceeds unity, which generically occurs as
the period is decreased.  

Over the last decade, 
the study of alternans \cite{FraSim,Couetal,Kar94,Vin,Pasetal,Foxetal,Quetal,Watetal},
and their control \cite{control}, has become a main
focus of research because of the potentially   
crucial link of this dynamical
instability with cardiac fibrillation \cite{Kar}.  
However, there is presently no simple analytical
understanding of how the bifurcation to alternans is manifested
\emph{spatiotemporally} in paced cardiac tissue.
Analytical progress to date is limited
to the one-dimensional circulation of electrical impulse in
a ring of tissue \cite{Couetal,Kar94,Vin}.

In this letter, we derive an
equation that governs the 
spatiotemporal dynamics of alternans close to the
onset of instability. This enables us to 
obtain a quantitative analytical understanding 
of the formation of recently observed
complex patterns of $APD$ oscillations that can promote
fibrillation \cite{Pasetal,Foxetal,Quetal,Watetal}.
A crucial feature of these patterns is that  
the $APD$ oscillates with opposite phases in two (or more) spatially extended 
regions of tissue, i.e. with a sequence LSLS...
in one region and SLSL... in the other.
These ``discordant alternans'' have been observed experimentally
in both two-dimensional \cite{Pasetal}
and linear strands \cite{Foxetal} of cardiac tissue,  
as well as in ionic model simulations \cite{Foxetal,Quetal,Watetal}. 
Moreover, they have been shown to lead to the formation
of conduction blocks \cite{Foxetal} as well as to the onset of spiral
wave formation and fibrillation \cite{Pasetal}. 
We show here that discordant alternans 
result from a \emph{pattern-forming} linear instability 
that has interesting similarities with classic  
instabilities leading to the spontaneous formation of
spatially periodic patterns in nature 
(such as Rayleigh-B\'enard convection, Taylor-Couette flow,
etc \cite{CroHoh}), but also presents some unique features.  

We consider  
a one-dimensional (1-d) homogeneous cable 
of length $L$ paced at period $\tau$ from one end ($x=0$). 
Close to the onset of instability, we can expand
the $APD$ and the period in the form
\begin{eqnarray}
APD^n(x)&\approx &APD_c\,+\,a(x,t)\,e^{i\pi n}, \label{aa}\\
T^n(x)&\approx &\tau_c\,-\delta \tau\,+\,b(x,t)\,e^{i\pi n},\label{bb}
\end{eqnarray}
where $APD_c$ and $\tau_c$ are the $APD$ and the period  
evaluated at the bifurcation point of the map 
($f'=1$), $x$ measures the position along the cable, and
$\delta\tau\equiv \tau_c-\tau\ll \tau_c$. 
In this range of period, $a$ and $b$
vary slowly from beat to beat, which allows us  
to treat  the time, $t\equiv n\tau$, as a continuous variable;
the fast beat-to-beat oscillations
are contained in the exponential factor 
$e^{i\pi n}$.  

A relation between $a$ and $b$ can first be
derived by noting that $T^n(x)$ is
the difference of arrival time of two subsequent
action potentials at $x$ \cite{Watetal}, or
\be
T^n(x)= \tau + \int_0^x \frac{d\,x'}{c(DI^n(x'))} -
\int_0^x \frac{d\,x'}{c(DI^{n-1}(x'))}, \label{disp}
\ee
where the first (second) integral on the right-hand-side  
is the time required for the leading front of the
action potential to
travel from the paced end to $x$ at the
$n^{th}$ ($n^{th}-1$) stimulus; concomitantly,
$c(DI)$ is the standard
dispersion curve that relates the propagation
speed of this front with the
local diastolic interval.
The dispersion curve is typically steeply
increasing at small $DI$ and flat at large $DI$. 
Substituting Eqs.
\ref{aa}-\ref{bb} in Eq. \ref{disp} with 
$DI^n(x)=T^n(x)-APD^n(x)$, and expanding to linear
order in $a$ and $b$, we obtain 
$b(x)\simeq \int_0^x a(x') dx'/\Lambda$ 
where we have defined $\Lambda\equiv  c^2/(2\,c')$, 
with $c$ and $c'\equiv dc/dDI$ evaluated 
at the bifurcation, and
assumed that $\Lambda$ is much larger
than the scale over which $a$ varies.

Next, in order to derive an evolution equation for 
the amplitude $a(x,t)$, we first neglect
the influence of the electrical coupling between cells on the
$APD$. This allows us to assume that the restitution 
relationship (\ref{res}), and hence the second iteration of the map
\begin{equation}
APD^{n+2}=f(T^{n+1}-f(T^n-APD^n)),\label{map2}
\end{equation}
continues to be valid even when
the $APD$ is non-spatially uniform; we shall soon see why it is crucial
to relax this assumption. 
We substitute Eqs. \ref{aa}-\ref{bb} in Eq. \ref{map2},
and expand the right-hand-side keeping only the dominant
linear and weakly nonlinear terms. Furthermore, we use the
aforementioned fact that, close to onset, $a$ varies slowly from beat 
to beat, and we therefore expand the left-hand-side as
$APD^{n+2}\simeq APD^n + 2\, \partial a/\partial n \, e^{i\pi n}$,
where $\partial a/\partial n=\tau\partial a/\partial t$.
Finally, we use the integral relation between  
$a$ and $b$ derived earlier. After equating both sides
of Eq. \ref{map2}, we obtain
\be
\tau \partial_t a =
\sigma a -g a^3 - \int_0^x \frac{dx'}{\Lambda}\,a(x')  ,\label{aeq}\\
\ee
where $\sigma\equiv f''(\tau - \tau_c)/2$, $g\equiv f''^2/4-f'''/6$, 
and all derivatives are evaluated at the bifurcation point.

In order to test this evolution equation,
we simulate the standard cable equation  
\be
\partial_t V = D\, \partial_x^2 V 
-\left(I_{\rm ion}+I_{ext}\right)/C_m,
\label{cable}
\ee
for the membrane current
$I_{\rm ion}$ given by the Noble model \cite{Nob} with
time in units of millisecond (ms),
$D=2.5 \times 10^{-4}$ cm$^2$/ms, $C_m=12 \,\mu$F/cm$^2$,
dx=0.01  cm, dt=0.05 ms, 
and $I_{ext}$  
modeling a sequence of stimuli applied   
at $x=0$ at the pacing interval $\tau$.  
To determine the parameters of the amplitude equation,
we compute the restitution and dispersion curves
by pacing Eq. \ref{cable} in a short cable and by using
two subsequent stimuli spaced by different
intervals to vary $DI$; we also use $V=-40$ mV as
threshold of the transmembrane voltage to define the $APD$.
We impose zero gradient boundary conditions on $V$ and
$a$ at the two ends of the cable in all 
our simulations.

\begin{figure}[t]
\centerline{
\epsfxsize=5.5cm\epsfbox{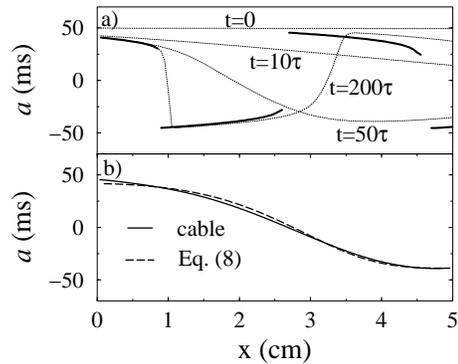}
}
\vspace{0.4 cm}
\caption{Amplitude $a$ of $APD$ oscillation vs $x$ for  
Noble parameters and $\tau=258$ ms.
(a): profiles obtained with Eq. \protect\ref{aeq}  
at different times (dotted lines) and
final stationary profile (solid line). (b): stationary
profiles obtained with Eq. \protect\ref{cable} (solid line)
and Eq. \protect\ref{aeqg} (dashed line). Nodes ($a=0$) separate
tissue regions with $\pi$ out of phase oscillations.
}
\label{comp}
\end{figure}

Fig. \ref{comp}(a) shows that Eq. \ref{aeq}
produces discordant alternans
consistent with the picture that  
restitution and dispersion suffice to produce this state
\cite{Foxetal,Quetal,Watetal}. 
However, the magnitude
of the spatial gradient of $a$ at the node  
increases with time, and can be shown to diverge in a finite time.  
This yields an unphysical spatial discontinuity of $APD$,
which also occurs if the system of coupled maps 
(Eqs. \ref{res} and \ref{disp}), from which the amplitude equation
is derived, is solved numerically. Note that the 
analogous system of coupled maps for 
a 1-d pulse circulating in a ring produces a smooth $APD$
modulation starting from a smooth initial condition \cite{Couetal},
which highlights the much stronger role of dispersion
at producing heterogeneity of $APD$ during pacing than circulation. 

This discontinuity, which is absent in the simulation
of the cable equation in Fig. \ref{comp}(b), can be cured
by adding spatial derivative terms 
to the amplitude equation. Since the 
cable is paced at one end, the underlying basic state (i.e. traveling pulses)
is not invariant under parity symmetry.
Hence, in addition to $\partial_x^2 a$,   
a term proportional to $\partial_x a$ must  
generally be included, which yields the final
form of our amplitude equation
\be 
 \tau \partial_t a = 
\sigma a -g a^3  
-\int_0^x \frac{dx'}{\Lambda}\,a(x')
- w \, \partial_x a+\xi^2 \partial^2_x a.\label{aeqg}
\label{aeqr}
\ee
To calculate the new lengthscales $w$ and $\xi$, we must
determine how the electrical coupling between cells
modifies the restitution
relationship (Eq. \ref{res}). For complex electrophysiological models
such as Noble, this generally needs to be done numerically using 
a procedure that will be discussed elsewhere.
For a simple two-variable
ionic model, which we study below, the analytical expressions
\begin{eqnarray}
w&=&2D/c,\label{w2v} \\
\xi &=&(D\times  APD_c)^{1/2}, \label{xi2v}
\end{eqnarray}
can be derived by interpreting Eq. \ref{cable} as a diffusion equation
with a source $I_{ion}$, and expressing
$V$ as space-time integral of $GI_{ion}$,
where $G$ is the standard Green's function of the diffusion equation.
This integral can then be evaluated analytically because the action potential  
shape is simply triangular for this model, and
used to calculate $w$ and $\xi$. Eq. \ref{xi2v} has 
the simple physical interpretation that $V$
diffuses a length $\sim \xi$ in the time interval of one 
$APD$. Therefore, the repolarization of a given cell
is influenced by other cells within a length $\sim \xi$ of cable.
In addition, repolarization of this cell
is influenced unequally by its left and right
neighboring cells because these cells are activated at different
times by the propagating wavefront. Clearly, this asymmetry must 
vanish in the limit $c\rightarrow \infty$ where all
cells are activated simultaneously consistent with Eq. \ref{w2v}.
Fig. \ref{comp}(b) 
shows that our regularized amplitude
equation (\ref{aeqg}) now produces a smoothly varying  
stationary profile of $a$ that agrees well with the 
simulation of the cable-Noble equation, where $a$  
is obtained from the $APD$ using Eq. \ref{aa}.

\begin{table}
\caption{Values of various lengths in cm with
$\lambda_{theor}$ (Eq. \protect\ref{lsta} 
or Eq. \protect\ref{ltra}) and $\lambda_{sim}$ (from simulations
of Eq. \protect\ref{cable}).}
\begin{tabular}{l||llllll}
 Model & $\Lambda$ & w & $\xi$ & $\lambda_{theor}/4$ & $\lambda_{sim}/4$ & $L_{min}$ 
 \\  \hline \
Noble &  49.1  & 0.045  & 0.18  & 2.33 & 2.6 & 2.75 \\
Two-variable &  3.55  &  0.031  &  0.235 & 1.33 & 1.1 &  1.15  
\label{lengths}
\end{tabular}
\end{table}

The genesis of discordant alternans can be understood  
by computing the linear stability spectrum
of the spatially homogeneous
state ($a=0$). We have calculated this 
spectrum both numerically for different $L$, 
and analytically for the large $L$ limit. The main result
is that the wave pattern can emerge from the amplification of
\emph{either} a unique finite wavelength 
mode, which yields a stationary pattern,  
\emph{or} from a discrete set of  
complex modes that approach a continuum in the 
limit $L\rightarrow \infty$, and yields a traveling pattern. 
There is indeed experimental evidence for both stationary
\cite{Pasetal} and traveling \cite{Foxetal} waves.

We treat here explicitly the large $L$ limit
since it provides the basis to understand  
finite-$L$ patterns. In this limit, we can analyze
stability by differentiating Eq. \ref{aeqg} with respect
to $x$ and letting $a(x,t) \sim e^{ikx+\Omega t}$,
with both $\Omega\equiv \Omega_r+i\Omega_i$  
and $k\equiv k_r+ik_i$ complex,
which yields at once the eigenvalue equation
\begin{equation}
\Omega=\sigma-\xi^2k^2-i\left[wk-1/(\Lambda k) \right].\label{spec}
\end{equation}
When dispersion is weak, 
a unique real mode with $k_i=0$ 
grows faster than the other complex modes 
and yields a stationary pattern. Its wavelength $\lambda=2\pi/k_r$,
is determined by the condition $\Omega_i=0$,
which yields  
\be 
\lambda = 2\pi (w\Lambda)^{1/2},\label{lsta}
\ee
in good agreement with the wavelength observed
in simulations of the cable-Noble equation
(Table \ref{lengths}). Note that $\cos k_rx$ is
an exact eigenvector of Eq. \ref{aeqg} linearized around
$a=0$ that satisfies  
$\partial_x a=0$ at the two cable ends when $L$ is
an integer multiple of $\lambda/2$.  
The threshold of instability occurs
when $\Omega_r=0$, or for a period $\tau_{th}$ 
defined by $\sigma_{th}=f''(\tau_{th}-\tau_c)/2=\xi^2/(w\Lambda)$.

In the opposite limit where dispersion is strong,  
complex modes that grow exponentially at 
large $x$ are the most unstable. It is simple to
deduce from Eq. \ref{spec} that each $k$-mode
travels towards the pacing end of the cable, but
a wave-packet constructed from a linear superposition of
these modes has a group velocity that makes the packet
move away from the pacing end. This is the signature of a 
convective instability \cite{CroHoh} where perturbations 
are transported as they grow, similarly to
Taylor-Couette vortices developing in an axial flow \cite{Babetal}.
In such a situation, patterns are only transient
unless the group velocity vanishes, 
or $\partial \Omega_i/\partial k_r=0$, and hence
they grow at a fixed point in space.
Moreover, the fastest growing wavelength that dominates at large time must
correspond to a maximum of $\Omega_r$, which yields the additional
condition $\partial \Omega_r/\partial k_r=0$. These two conditions
are equivalent to requiring that $d\Omega/dk=0$. From this, 
we deduce that the threshold of absolute instability occurs
when $\sigma_{th}=(3/2)(\xi/2\Lambda)^{2/3}$, with a pattern of wavelength
\begin{equation}
\lambda=(4\pi/\sqrt{3})(2\xi^2 \Lambda)^{1/3},\label{ltra}
\end{equation}
which travels with phase velocity $\Omega_i\lambda/(2\pi)$
where $\Omega_i=(3\sqrt{3}/2)(\xi/2\Lambda)^{2/3}$. 
It can also be deduced that
traveling waves are favored over
a stationary pattern when $\Lambda=2c^2/c' \ll \xi^4/w^3$,
and hence for large enough dispersion ($c'$),
and vice versa in the opposite limit.

Nonlinearities are mainly 
responsible for saturating the amplitude of the linear
modes, but both the wavelength and evolution of the 
pattern can generally vary with distance from onset.
There is a close analogy between the amplitude equation
(\ref{aeqg}) and the real Ginzburg-Landau equation that has been
extensively studied in the context of phase transitions and
front propagation \cite{CroHoh}. The dynamics is richer here because the 
integral term originating from dispersion 
causes a non-local interaction of
the fronts separating two out-of-phase oscillating
regions with the pacing end of the cable.

\begin{figure}[t]
\centerline{
\epsfxsize=6.6cm\epsfbox{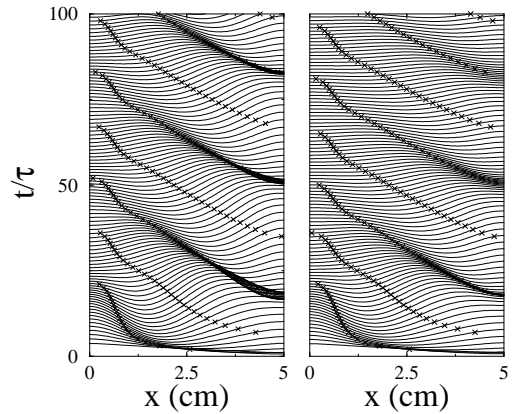}
}
\caption{Space-time plots of $a$  
obtained by simulations of Eq. \protect\ref{aeqg} for
parameters of the two-variable model, showing   
absolutely unstable (left and $\tau=295$ ms) and  
convectively unstable (right and $\tau=298$ ms ) wave patterns. The crosses denote
the positions of the nodes ($a=0$).
}
\label{examples}
\end{figure}

As a final test of our theory, we study
a two-variable model where all the parameters
of the amplitude equation can be calculated analytically
including $w$ and $\xi$ given by Eqs. \ref{w2v}-\ref{xi2v}, and  
for which our theory predicts traveling waves.
In this model, $I_{\rm ion}/C_m$ is the
sum of a slow outward current, $I_o/C_m=\tau_0^{-1}(S+(1-S)V/V_c)$,
and a fast inward current, $I_i/C_m=-\tau_a^{-1} h \,S$, where inactivation 
of the latter is controlled by  
\be
\partial_t h=(1-S-h)/(\tau^-(1-S)+S\tau^+)
\ee
In addition, $V$ is dimensionless
and $S\equiv (1+\tanh((V-V_c)/\epsilon))/2$. We choose
$V_c=0.1$, $\tau_0=150$ ms, $\tau_a=6$ ms, $\tau^-=60$ ms, 
$\tau^+=12$ ms, $\epsilon=0.005$, and we simulate Eq. \ref{cable} with
the same $D$ as before, dx=0.01 cm, dt=0.02 ms, and $V=0.1$ as threshold  
for the $APD$.  

\begin{figure}[t]
\centerline{
\epsfxsize=9cm\epsfbox{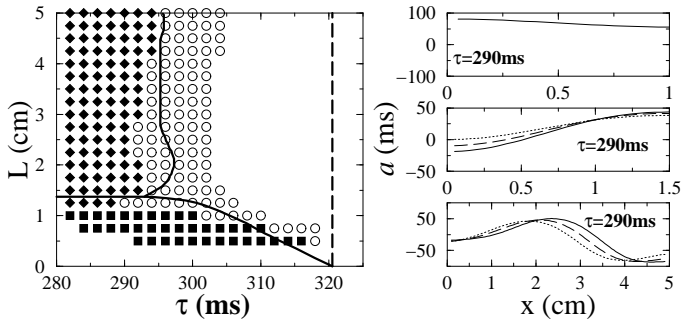}
}
\vspace{0.2 cm}
\caption{Stability diagram of
two-variable cable model with domains of no-alternans
(open circles), concordant alternans
(filled squares), and discordant alternans
(filled diamonds); conduction blocks form
at smaller $\tau$ not shown here. Boundaries between the 
same domains obtained by simulations
of the amplitude equation (\protect\ref{aeqg})
are shown by solid lines. The dashed line denotes the bifurcation period
for alternans predicted by the map of Eq. \ref{res}.
On the right panel we show profiles 
of $a$ vs $x$ at times $t$ (solid), $t+2\tau$ (dashed) and $t+4\tau$ 
(dotted lines), for three different cable lengths.}
\label{phasediag}
\end{figure}

As predicted, we observe traveling waves in this model
with a wavelength that agrees well with Eq. \ref{ltra}. 
Fig. \ref{examples} illustrates
sustained and transient wave patterns above (left) and
below (right) the onset of absolute stability, respectively. 
Furthermore, Fig. \ref{phasediag} shows that
the results of simulations of the cable and amplitude
equations are in good agreement over a wide
range of $L$ and $\tau$, including the
fact that the onset of instability occurs at a shorter
period in a larger cable (and, hence, for a slope of the
restitution curve larger than unity).
Fig. \ref{phasediag} is qualitatively similar
for stationary waves, but this 
overstabilization is smaller because dispersion is weaker.  

The $APD$ oscillations  
at a fixed $x$ are periodic when the
wave is stationary and quasiperiodic when it travels. 
For both cases, we find here that 
$\lambda$ is independent of $L$, in contrast to
the oscillations produced by a pulse circulating
in a ring where it is known that
$\lambda \simeq 2L/(2i+1)$ for weak dispersion  
\cite{Couetal} (with $i$ integer and $L=$ ring perimeter). 
As will be discussed elsewhere,
our theory applied to the ring shows that  
the bifurcation to alternans is finite dimensional
with $i=0$ being the most unstable mode, in agreement with
the fact that it is this mode that is generically
selected in experiments \cite{FraSim} and
ionic model simulations \cite{Couetal,Kar94,Vin}.  
In addition, it
shows that the 
gradient term ($-w\partial_x a$) can lead to
quasiperiodicity even in the absence 
of dispersion ($c'=0$).

Our results demonstrate that the formation
of discordant alternans is crucially affected by the 
effect of electrical coupling (diffusion) on repolarization,
in addition to restitution and dispersion. 
Dispersion is responsible for the formation of nodes and
spatial gradients of $APD$ that steepen with time. 
Diffusion, in turn, tends to spread the $APD$ spatially,
and also induces a drift of the pattern away from the pacing site
that is induced by the more subtle gradient term 
($-w\partial_x a$) in the amplitude equation. When
dispersion is sufficiently weak, drift balances dispersion
and produce a stationary pattern. In the opposite limit,
the tendency for dispersion to form steep gradients
of $APD$ is balanced by the spreading effect of diffusion. 
Nodes then travel, cyclically 
disappearing (appearing) at the pacing (opposite)
end of the cable. 

In conclusion, we have derived a simple  
evolution equation that describes the universal spatiotemporal
dynamics of small amplitude alternans in paced cardiac tissue. 
Moreover, we have shown that discordant 
wave patterns \cite{Pasetal,Foxetal,Quetal,Watetal}, 
which are linked to fibrillation \cite{Pasetal}, 
result from a finite wavelength
linear instability. Hence, their formation
requires a minimum tissue size 
$L_{min}\sim \lambda/4$, required for at least
one node to form. The value of $L_{min}$ that we
measure in simulations of reaction-diffusion
models are actually close to $\lambda/4$
with $\lambda$ predicted by Eq. \ref{lsta}
and Eq. \ref{ltra}, respectively 
(Table \ref{lengths}).  
This equation can be extended
to higher dimensions as well as to include 
heterogeneities. Our preliminary results in a 2-d
homogeneous tissue are very similar to 1-d.
Finally, this equation is readily applicable to model  
a non-constant pacing interval and provides
a theoretical basis to study the control of 
alternans in spatially extended tissue.

This research was supported in part by NIH 
SCOR in Sudden Cardiac Death P50-HL52319.

\end{document}